\documentclass[cup5b]{caps}
\usepackage{graphicx}
\usepackage{amssymb}
\usepackage{ociwsymp1}   

\HeadText{S. M. Faber}  

\begin{document}

\pagenumbering{arabic}

\author[]{S. M. FABER\\UCO/Lick Observatory, University of California, Santa Cruz}

%
%

\chapter{CIW Cosmology Symposium: Conference Summary --- Observations}

\begin{abstract}

I review the major concordances and controversies of the meeting
concerning observations.
Cosmologists are nearing agreement on the global cosmological parameters
of the Universe.  A few parameters still have large error
bars, notably $\Omega_{mat}$, but most of these windows will close
soon with upcoming accurate observations of the CMB.  The major era of
chasing the major cosmological parameters is now closing.  Understanding
galaxy formation in all its messy detail will continue to occupy cosmologists for 
some time to come.  Development of highly accurate standard candles,
such as inspiraling black holes, holds out hope for understanding $\Lambda$,
whether it is truly constant or evolving.  This is the major prospective
contribution from cosmology to fundamental physics in the next generation.
I close with a discussion of anthropic cosmology.  Anthropic reasoning
has been shown to be correct at least three times in the history
of cosmology.  Applying it now leads us to take seriously the
prospect of other universes, a notion that should be pursued seriously
by theoreticians.

\end{abstract}

\section{Introduction}

     If the birth of cosmology can be reckoned from the first 
in-depth survey of the Universe 
beyond our Galaxy, that birth dates to the beginning of the 20th century, when the 
first comprehensive photographic survey of external galaxies was compiled by Keeler at Lick.  
Dated thus, cosmology has reached its centennial milestone, coinciding providentially 
with the centennial milestone of the Carnegie Institution of Washington.  At this 
conference we therefore celebrate not one but two birthdays, and it is appropriate to 
appraise the progress in our field after a century of effort.

     The efficiency of the human scientific endeavor is such  
that one hundred years of work in a 
field typically yield enormous fruit, and cosmology is no exception.  Given the 
remoteness of the object of our study in both distance and time, what we have learned in 
100 years is nothing short of stupendous.  The basic outlines of the subject are now 
known, the fundamental questions have been framed, and many have even been 
answered.  We now have a theory for the global geometry of space-time, we 
have a general description of the time evolution of the system from inflation to now, 
and we are closing in on the values of the dozen or so global parameters that are needed to 
characterize the observable Universe.  We have mapped the heavens and seen structure, and we 
have a rough theory for why that structure exists and how it developed.  In short, the state 
of our field can accurately be characterized as ``mature.''  

     In this ocean of knowledge, three islands of ignorance still stand out.  
The first is 
the messy baryonic physics of galaxy formation, which is challenging but not in any sense 
fundamental.  It will keep us busy for another couple of decades 
but will ultimately yield to the heavy artillery of future high-powered computing.  
There is no fundamentally new physics to be discovered here.  

     The second island is the domain of the question, what came ``before'' the Big Bang, 
and indeed, whether we can presently frame any meaningful question 
at all along these lines.  The ``cause'' of the Big Bang takes us to the very 
depths of epistemology and natural philosophy.  I shall have a few words to say about 
this at the end of this talk, but the basic conclusion is that, as scientists, we do not appear 
ready to grapple meaningfully with this mother of all questions at the present time.

     The third island, by virtue of its in-between status, is the most exciting for exploration 
now.  This is the realm of the {\it dark energy}.  Dark energy is not so far 
removed from established principles of particle physics that it is unapproachable 
by known methods, and it is also accessible to measurement via observations of the 
distant Universe. Its pursuit promises to enlarge our basic concept of ``thing'' by 
providing an entirely new thing to consider in addition to radiation and matter, and in the 
process it may cause us to examine what other new ``things'' can in principle exist.  It 
also compels us to question the future evolution of our Universe and whether new 
physics might appear in the future that will govern its ultimate fate (I am here 
invoking an analogy to early inflation, whose physics was 
the mother of our present-day Universe).  We are very lucky, I think, to have discovered 
this third land, as cosmology would otherwise be confined today to the challenging yet
basically trivial pursuit of galaxy formation, versus the fundamentally 
impenetrable and inapproachable realm of the Big Bang.  Dark energy provides
something meaty yet doable in between and
promises to keep cosmology intellectually vibrant for some time.  
     
     I was amused at this conference at how some of us evidently feel discomfited by 
recent success.  We are closing in not only on broad concepts but also on specific details, 
yet some attendees felt (almost reflexively?) compelled to express doubt.  Consider, for 
example, Virginia Trimble, who said pessimistically, ``I wouldn't bet the window on 
$\Omega_{mat}$ to be any smaller than 0.1 to 0.5 with 95\% confidence.''  Or Malcolm Longair, 
who anguished whether ``we might be extrapolating too far...,'' ``there might be 
surprises..,'' or ``fundamental misconceptions'' in the present picture.   Or Bernard 
Sadoulet, who wondered whether the discovery of $\Lambda$ might even signal ``the first 
hint of a failure of gravity.''  I began to wonder myself whether WIMPS had actually 
been detected at this conference for the first time!

     But MACHOS were also here in quantity. Mike Turner exuded typical
confidence with 
statements like: ``There are no current controversies,'' and ``Now that we're closing 
in.''  Andreas Albrecht advised us to ``stop whining [over small discrepancies]
and get to work!''  I personally 
identify more with the MACHOs --- the data are fitting so well together (by virtue of non-zero
$\Lambda$) that it seems highly unlikely that the whole edifice could be 
seriously undermined at this point.  Musings about ``missing something big'' strike me as 
wishful thinking that the g Primack, J. R.rand game of chasing cosmological parameters 
might continue indefinitely, when in fact a major 
era is closing.

     Tactics in this end game are shifting, reflecting the new situation.  When a science is 
young, a single observer with a simple apparatus can make a paradigm-shifting 
discovery.  Thus, we cosmologists used to go to the telescope with the goal of measuring 
a single number, like $H_0$ or $q_0$.  No more.  With the possible exception of $H_0$, 
hardly any experiments measure a single number any more, but rather increasingly 
complicated functions of the form F($h$, $\Omega_{tot}$, $\Omega_{mat}$, $\Omega_{bary}$, 
$\Omega_{\Lambda}$, $\sigma_8$, $b$,...).  Where these surfaces intersect is the sweet spot of 
the Universe.  Each experiment in this new era has only a piece of the truth, and   
this has led to a new era of coordination, cooperation, 
and intercomparison, but the 
old satisfaction of pioneering on alone is somewhat diminished.  

\section{Current Status and Prognostications}

     That said, recent progress has been outstanding, and I list here five major 
experiments that all  of us would agree have provided big breakthroughs. 

\medskip
$\bullet$ {\it The $H_0$ Key Project} value $72\pm8$ (Freedman,
Jensen) came just in time to anchor measurements of other 
cosmic parameters from CMB experiments and redshift surveys.

$\bullet$ {\it CMB surveys} (BOOMERANG, MAXIMA, DASI, CBI; Wright) confirmed
the flat universe predicted by inflation, and the rest of the CMB spectrum 
tightly constrained other 
combinations of important cosmic parameters.

$\bullet$ {\it Type Ia supernovae} (Filippenko, Perlmutter) directly 
detected $\Omega_{\Lambda} \sim 0.7$, exactly the right value to fill
the empty gap between $\Omega_{tot} = 1$ from 
the CMB and $\Omega_{mat} \sim 0.3$ from LSS and dynamics.
Cosmologists might have 
wrangled endlessly about the reality of an $\Omega_{\Lambda}$ deduced arithmetically
from $\Omega_{tot}$ and $\Omega_{mat}$, 
but the direct detection from Type Ia's has virtually put that controversy to rest.

$\bullet$ {\it  Big Bang nucleosynthesis} (Tytler, Steigman) yielded 
the first model-independent estimate of $\Omega_{bary}$, from primordial deuterium.  

$\bullet$ {\it  Massive nearby redshift surveys (2dF and SDSS)} (Colless, Bernardi)
 tightened the noose 
on large-scale structure, constrained $\sigma_8$ 
and bias, $b$, and  demonstrated the power of huge 
samples for studying structure formation.  Bringing the fire-power 
of these surveys to
bear on galaxy formation will be the logical next step.

\medskip
     From this cascade of  recent data, certain fundamental
conclusions have emerged:

\medskip
$\bullet$ The Universe is essentially flat: $\Omega_{tot}$ = 1, as predicted by inflation. 

$\bullet$ Dark matter exists and is mainly non-baryonic.  Taking $\Omega_{bary} = 
0.04 h_{70}^{-2}$ and $\Omega_{mat}$ = 0.15--0.35 (see below), we find 
$\Omega_{tot}/\Omega_{bary}$ in the range 4-8, with the value of 1 totally excluded.

$\bullet$ $\Omega_{\Lambda}$ is non-zero and in the range 0.65-0.85.  This is implied 
indirectly by the previous two bullets but also comes directly, as I have noted, 
from Type Ia supernovae.

$\bullet$ Primordial fluctuations created in an early epoch of inflation planted the 
seeds for later formation of galaxies and large-scale structure via gravitational instability.

\medskip
    In contrast, discussion at this conference shows that 
several important parameters still remain insecure:

\medskip
(1) The total window on $\Omega_{mat}$ quoted at this 
meeting was 0.15--0.35, which is still quite large.  
Neta Bahcall marshaled an impressive case for 
0.2, but many of the methods she quoted are fairly model
dependent. Values of $\Omega_{mat}$ from peculiar motions, 
large-scale structure, and the CMB in contrast tended to hover near or above 
0.3 (Dekel, Colless, 
Bernardi).  The best strategy is to wait, as the most accurate 
value will come from the next round of CMB measurements (WMAP).

(2) Similar controversy swirled around $\sigma_8$, which is
expected because most methods measure the product 
$\sigma_8 \Omega_{mat}^{0.5-0.6}$, so when one goes up the other 
goes down.  Quoted values of $\sigma_8$ ranged from 0.7 to 1.0, with Bahcall 
favoring the higher range and Colless the lower.  New data from the Cosmic Background 
Imager (Readhead) on short wavelengths also favors $\sigma_8 \sim 1$.  
In retrospect, the $\sigma_8$ scale is a bad choice 
for normalizing the power spectrum because it
sits between Ly-$\alpha$ and large-scale structure surveys, so
that quoted values from these data sets are 
often entangled with the assumed spectral index, $n$. 
The final value of $\sigma_8$ will require 
a joint analysis of the entire body of fluctuation data (CMB, LSS, Ly-$\alpha$), 
which again awaits the next round of CMB data.


(3)  We still do not understand the density run of 
matter in galaxies, and uncertainty here generated a 
modest rear-guard attack on the Key Project value of
$H_0$.  Broadly,  dark-matter halo models 
(e.g., Nararro, Frenk \& White 1997) predict a fairly shallow dark-matter mass profile
on the scale of galaxy-galaxy lensing, $\rho \propto r^{-\eta}$, where $\eta$
is about 2. This is consistent with gravitational lensing time delays only if $H_0 
= 50$ (Kochanek).   For power-law mass models, $H_0 \sim (\eta-1)$; hence, to 
permit the Key Project value of $H_0 = 72$ would
require $\eta \sim 2.5$, which is already that of the light alone; adding dark
matter would only make this shallower, reducing $H_0$ below 72.
At smaller radii the picture is also confused.  Here, adiabatic
contraction of dark matter by baryonic infall should 
produce rather steep central slopes and densities (Blumenthal et al.~1987),
yet a variety of information points to the contrary, including
inner galaxy rotation curves (e.g., de Blok \& Bosma 2002, Swaters et al.~2003),
lensing around central cluster galaxies (Ellis), 
rotation curve amplitudes (Alam et al.~2002), the amplitude of
central bars (Weiner, poster paper), and the
scarcity of dwarf galaxies (Dekel).

The disconnect between theory and observations for $\rho(r)$ in halos has led some
to posit warm or self-interacting dark matter (reviewed by Silk).  Alternatively, halo
models may need revision to include stirring of dark-matter cusps via
bars or mergers (Katz), strong feedback and consequent ejection of baryons (Silk, Dekel), 
photoionization to retard baryonic infall (Katz), and
stripping (Katz).  A third possibility, not much discussed here, is a tilt
to the fluctuation spectrum, $n$, to reduce the strength
of fluctuations on galactic scales, though that might imply late structure
formation and consequently galaxy formation too late to match
the number counts of LBGs and early QSOs (Somerville, Primack \& Faber 2001).  
My guess is that the $\rho(r)$
discrepancy will be 
resolved by a combination of several small 
things, such as small changes to the lensing time delays (which are  
accurate to only $\sim$15\%; Chartas, Treu), improvements to galaxy collapse models (as
noted above), a small reduction in small-scale fluctuation amplitudes via spectral tilt, 
and a small downward adjustment of $H_0$ to the mid-60's, 
still within the current Key Project window.
Should future data confirm the high value of $H_0=72$ from the Key Project, this will
exacerbate the tension between model slopes and density data, motivating further
 scrutiny of the collapse models.

(4) The above discussion makes clear that 
$H_0$ not quite as rock-solid as we would wish, and so it is worth reviewing the 
values of $H_0$ discussed at this conference.  
In addition to the revised Key Project value of $72\pm8$ 
(Freedman, Jensen), four other independent values were mentioned.
The combination of the CMB plus nearby large-scale structure is consistent with
$H_0 \sim$ 70 (Colless).  The CMB itself implies that $H_0 \sim 70$ if the Universe
is perfectly flat.  The maser galaxy NGC 4258 implies an upward 
correction of 10\% to the Key Project value to 80 or so (Freedman), 
but the photometry is not HST's 
best and should be redone.  Finally, there are gravitational lenses, 
which, if $\eta$ really is --2.5, 
imply a downward correction to $H_0$ = 50 (Kochanek).  
Several speakers lamented the shaky distance 
to the Large Magellanic Cloud, which underpins the Key Project value.  Some also 
mentioned a downward correction as large as 10\% to the Key Project value 
owing to our location in a local low-density bubble.  My own guess, $H_0=65$, is 
a compromise reached by stretching the error bars of all methods to 
achieve overlap.  Finally, if cosmologists are willing simply to {\it posit} that the Universe
is exactly flat, CMB data will determine the value of $H_0$ to
exquisite accuracy, and that might be in fact be our final adopted value.

(5) David Tytler's review of  $\Omega_{bary}$ from 
BB deuterium  was illuminating yet a bit scary.  The number of QSOs that has
been analyzed is only 5, of which only 2 are really firm.  Steigman stressed that
there are residual discrepancies for both He and Li obtained when using the $\Omega_{bary}$
measured from deuterium.  Given the scatter and small numbers, I am concerned that the data 
presently do 
not merit an error bar as small as the 10\% that is often quoted.  Again, this question will
be resolved soon when the second peak in the CMB spectrum is accurately measured.

     The following table is a playful attempt to predict how some
of these issues will be resolved in future:

\medskip
\hangindent=25pt\noindent
2006 $H_0 = 65\pm3$ is determined by using a combination of the CMB and LSS.

\hangindent=25pt\noindent
2008  Dark matter is detected in the lab as the LSP; somebody (Bernard Sadoulet?) wins 
the Nobel prize.

\hangindent=25pt\noindent
2008 GAIA claims to measure $H_0 = 75\pm3$ by providing accurate parallaxes
for a few cepheids but 
cosmologists take no notice; the distance-ladder approach is by then deemed dead.

\hangindent=25pt\noindent
2010 Moore's law rescues ``gastrophysics.''  The cusp/concentration problem with 
baryons and dark matter in galaxies is resolved via 
a bit of everything: fewer baryons falling into galaxies, stirring of central 
cusps by bars and mergers, expulsion of gas (especially in dwarfs) by winds, etc.

\hangindent=25pt\noindent
2012  LISA measures $w$=$-$1.01$\pm0.01$; $w^{\prime}$ = 0.03$\pm$0.05 using the 
newly developed ``ultimate'' standard candle, inspiraling central black holes (Phinney).  
Confronted finally by what appears to be a genuine cosmological constant, string theorists
retire in droves.

\hangindent=25pt\noindent
2015  The fourth and final reanalysis of  PLANCK data teases out the tensor B 
modes (Cooray, Zaldarriaga)...but barely at the 2-$\sigma$ level of significance.  
Undeterred, NASA trumpets to Congress, ``Now 
we've {\it really} seen the face of God.''

\hangindent=25pt\noindent
2030 Neal Katz retires, having finally removed all but one free parameter from 
galaxy formation models.  The one remaining remaining free parameter is, of course, the 
star-formation rate.

\medskip\noindent
The fifth entry concerning inspiraling black holes is particularly interesting.  If these
objects can be made into few-percent standard candles, as Sterl Phinney
outlined, they hold the prospect for actually measuring $w^{\prime}$, and
thus testing whether $\Lambda$ is indeed constant or evolving.

\section{Anthropic Cosmology}

     I close this summary by coming out 
of the closet as a believer in anthropic cosmology.  Much has been 
written on this topic, both pro and con, some of it needlessly
complicated, and I'd like to take this opportunity to state my 
view, since I envision anthropic reasoning to play a greater role in cosmological 
discussions in the future.  Anthropic arguments are
a kind of data, though not of the conventional kind.

     To illustrate, consider the plight of an intelligent cosmologist back in the geocentric 
Aristotelian era.  In the then-current world model, the radius of the Earth would have 
looked like a fundamental constant of the Universe, 
analogous to the radius of today's Universe.  Thinking 
like a modern cosmologist, our Greek would have felt compelled to 
understand where this value, the radius of the Earth, came from.  There would have been 
two choices: search for an argument rooted in physical principles that shows 
why this radius could have one and only one conceivable value, namely that observed. 
Or, posit the existence of an infinite (or at least very large) ensemble of 
spherical bodies 
with a spectrum of radii, and then argue that the actual Earth must occupy a narrow 
window within that spectrum that is picked out {\it a posteriori} 
by the existence of life as we know it on this Earth 
(the precise location of the radius within this window would be random.)  The first route might 
be termed the ``physics approach.''  I argue that the situation of the Greek is not 
fundamentally different than the situation today with modern cosmology, 
and that the physics route, if followed 2000 years ago, would have
been demonstrably sterile.  The 
correct approach for the Greek would in fact be the anthropic approach.  

    So far, this is familiar, but I now make three points that have not 
been stressed previously.  First, cosmology has 
been faced with many explanatory challenges in the past similar to the ones we now face, 
and in all cases the correct approach was in fact anthropic.  Classic 
examples are the nature of the Sun, the nature of the Solar System, and the nature of the 
Galaxy.  We now see that there is nothing unique about any of these objects and that 
their properties were in fact picked out of a much larger ensemble by anthropic 
requirements.  Thus, anthropic arguments are not speculative, they have been proven correct 
several times over.

     The second point is that an anthropic argument makes sense only if you accept the 
actual existence of the larger ensemble --- {\it even if you have not yet observed it.}  
The larger ensemble is not merely hypothetical, it is really out there!  This is the real power of 
an anthropic argument --- {\it not} to explain a particular cosmological parameter but 
to alert us in to the existence of a much larger (though possibly still unseen) 
ensemble.  Again, this is borne out by history.  In the above cases,
astronomers actually went on to discover semi-infinite ensembles of 
suns and galaxies, and we are now (2000 years later) well on our way to 
verifying a semi-infinite ensemble of planets.  In all cases, 
the singular object that we were once so fixated on was revealed to be merely
a member of a 
much larger sample.

     The third point is that, once the anthropic approach is invoked, the thrust of the 
science shifts from trying to explain the properties of the singular object  
to understanding the {\it properties of the ensemble.}  Not, why is our Sun the 
way it is, but rather, what does the total ensemble of star-like objects look like?  What are 
the physics of stars in general, how do they form, what is the range of properties spanned 
by the class as a whole, and how does one characterize a given star within the class?  
Again, this has been the standard route of astronomical inquiry, which has borne 
tremendous fruit.
 
      In our present situation as cosmologists, I argue that taking an anthropic 
approach to explaining the properties of our Universe is a rational strategy 
based on historical success.  This presumably 
includes not only the dozen or so macroscopic
cosmological 
parameters cataloged by Freedman in her review (this conference), but also all the messy 
40-odd parameters of the particle physics
Standard Model (however they might ultimately be revised).  
Together these parameters 
are simply the suite of numbers needed to characterize our Universe.  
Anthropic reasoning then leads us to accept, or at least hypothesize, 
the existence of the larger ensemble, namely {\it 
other} universes.  It would be reassuring  as we do so to have at least a glimmer of a 
physical process that might have created that ensemble.  Fortunately, the speculative 
fringes of particle physics and cosmology have come up with a few ideas revolving around 
chaotic inflation and multiple dimensions.  Unfortunately, there does not seem to be the prospect 
of directly observing these other universes any time soon.  Nevertheless, that 
should not stop us from trying to deduce their properties,
any more than the Greek cosmologist should have been deterred from thinking
about other planets.  His attempts might have failed for lack
of proper tools and understanding at the time, but they could not in any sense have
been termed ``unscientific.''

     My last point is a slight stepping back from a purely anthropic approach and is 
occasioned by the fact that at least one cosmological parameter, $\Omega_{tot}$, 
{\it has} been explained by appeal to fundamental physical
principles. The fundamental process generating $\Omega_{tot}$ is inflation, 
which in its simplest version produces a very flat universe.  
Inflation, in turn, is believed to occur under a wide range of
conditions, and is thus regarded by particle physicists as 
at least generic, if not ubiquitous.
The case of $\Omega_{tot}$ thus warns us that perhaps not
all current cosmological parameters are equal, in the sense that some may 
one day be derived 
from others via generic physical arguments.  To be more precise, I am reasoning 
here that the capacity of our Universe 
to support inflation (and hence have $\Omega_{tot} \approx 1$)
was determined by a combination of 
other, more fundamental cosmological parameters.  If this reasoning is true, it suggests 
that a continuing 
job of cosmologists will be to discover hidden relationships among the current 
parameters based on physical principles, and in the process to reduce the total number of 
independent parameters.  This activity will resemble
conventional physics, i.e., explaining one thing by another. 
However, the history of astronomy and cosmology strongly suggests that even diligent 
application of this method will leave some number of cosmological
parameters ultimately unaccounted for.  The leftover ones will 
be the truly fundamental ones and thus, I argue, will have to be accounted for anthropically. 
The frontier will shift to discovering the 
properties of the larger ensemble spanned by these remaining
fundamental parameters.  Not, why is our Universe 
the way it is, but rather, what does the greater ensemble of universes, the {\it 
Meta-universe}, look like?

     In closing, I would be the first to
admit that the anthropic explanation for our Universe does not provide 
the ultimate explanation for the Universe that we are looking for.  
It does not address the origin of the Meta-universe, and hence 
postpones by only one step the reckoning of ultimate causes --- 
why something exists rather than nothing, and indeed whether that question
has meaning.  Perhaps the Theory of Everything, if it lives up to its
name, will explain the existence and properties of the Meta-universe from
first principles.  Until then, I find it thrilling to contemplate the possibility
of multiple universes in parallel with our own, and count that awareness a major
step forward in the growing cosmological sophistication of our species.

     There is a convention in astronomy (not strictly observed but still useful) that
capitalizes the name of an object when it refers to our local example, as distinct from an 
object in the larger ensemble.  Thus, ``earth'' becomes ``Earth,'' ``sun becomes ``Sun,'' and
``galaxy'' becomes the ``Galaxy.''  In keeping with this tradition, I suggest that
anthropic cosmologists might capitalize the word ``Universe'' when 
referring to our own, to express explicitly our willingness to contemplate 
the existence of the larger 
ensemble.


\begin{thereferences}{}

\bibitem{}
Alam, S. M. K.,  Bullock, J. S. \& Weinberg, D. H. 2002, ApJ, 572, 34

\bibitem{}
Blumenthal, G. R., Faber, S. M., Flores, R., \& Primack, J. R. 1986, ApJ, 301, 27

\bibitem{}
De Blok, W. J. G., Bosma, A., 2002, A\&A, 385, 816

\bibitem{}
Navarro, J. F., Frenk. C. S., \& White, S. D .M. 1997, ApJ, 490, 493

\bibitem{}
Somerville, R. S.,  Primack, J. R. \& Faber, S. M. 2001, MN, 320, 504

\bibitem{}
Swaters, R. A., Madore, B. F., Bosch, F. C. \& Balcells, M. 2003, 
ApJ, 583, 732

\end{thereferences}
\end{document}